# Designing Interactions with Furniture
## *Towards Multi-Sensorial Interaction Design Processes for Interactive Furniture*

Pedro Campos[1], Nils Ehrenberg[1] and Miguel Campos[2]

[1]*Madeira-ITI, University of Madeira, Campus da Penteada, Funchal, Portugal*
[2]*WowSystems LLC, R. Mary Jane Wilson, 21 G, Funchal, Portugal*
{pedro.campos, nils.ehrenberg}@m-iti.org, miguel.campos@wowsystems.pt

Keywords: Human-Computer Interaction, Usability, Interactive Furniture, User Experience, Human Factors, Ergonomics, Embedded Systems, Interaction Design.

Abstract: In this paper, we argue for novel user experience design methods, in the context of reimagining ergonomics of interactive furniture. There is a need for improving both creativity and productivity at the workplace, and there is ample room for scientific advancements brought by embedded systems, sensors and actuators which can now be part of future pieces of furniture. Creative industries' workers are becoming more prominent as countries move towards intellectual-based economies. Consequently, the workplace needs to be reconfigured so that creativity and productivity can be better promoted at these spaces. This position paper presents several directions that can shed light on how we can better design interactive furniture for the workplace. In particular, we argue for a multisensorial approach as a promising way of achieving the above-mentioned goals.

## 1 INTRODUCTION

Creative industries and their workers are becoming more and more prominent, as countries move towards intellectual-based economies. Consequently, the workplace needs to be reconfigured so that creativity and productivity can be better promoted at these spaces. This position paper introduces the design and prototyping of a new seat for relaxation and for productivity, which fuses the digital and the analog in a futuristic piece of furniture, targeted at workspaces, in particular creative co-working spaces. This interactive seat, with embedded sensors and actuators, aims at reimagining the role of interactive technologies in the workspace of the future, especially by challenging disruptions in shared workspaces. Our project aims to explore the balance between social interruptions and social benefits of the creative workspace. Creative industries' workers are becoming more prominent as countries move towards intellectual-based economies. Consequently, the nature and essence of the workplace needs to be reconfigured so that creativity and productivity can be better promoted at these spaces. We introduce initial interview studies, a co-creation workshop - both of which informed the design - and early renders and 3D prints of our prototype.

We also present the rational for how the design has been developed and our initial findings from the prototyping efforts. There is a significant need for improving both creativity and productivity at the workplace, and there is ample room for scientific advancements brought by embedded systems, sensors and actuators which can now be part of future pieces of furniture. We will start by describing the most important human factors in this context of design (Section 2, Human Factors in the Workplace); afterwards, we describe our design approach (Section 3), including the collaborative prototyping sessions and resulting artefacts; finally, we conclude with our position regarding the need towards multisensorial approaches for designing interactive furniture and some ways this can be achieved.

## 2 HUMAN FACTORS IN THE WORKPLACE

There are positive correlations between creativity-supporting work environments and product innovation (Dul and Ceylan, 2014). Organizations therefore seek to engineer their workspaces in order to better support creativity through ergonomics, e.g. by including physical elements that can systematically improve the employee creativity levels (Dul and Ceylan, 2011). Productivity can be defined as the effectiveness of converting effort into useful outputs. In general, organizations seek to improve their productivity because it is a critical determinant of cost efficiency and better outcomes. Our approach is based on reimagining office furniture and designing it in such a way it becomes a place to relax, to regain focus and to conduct creative work. We are currently conducting three pilot user studies at co-working spaces in three different locations: Malmo and Lisbon. In parallel, we have been prototyping in 3D (3D-image renders as well as a 3D-printed physical prototype) a new seat/workstation for improving creativity at work. In this position paper, we argue that any design approach for interactive furniture should be grounded in informed studies and user observations. We divide our analysis between the physical factors at the work environment, and the disruptions that happen during quotidian work.

### 2.1 Physical Work Environment

We first focus on the physical workspace aspects affecting creativity, proposing the creation of an open pod-like workstation unit that can help creativity by simulating a creativity-supporting work environment. Typical physical environment improvements, that affect employee's creativity in a positive way, as suggested by various researchers, include: a non-crowded workspace, the presence of plants, the use of inspiring colors on the walls, a new carpet in the office, more pictures and posters on the walls, windows with an outside view, privacy, dim lightning, etc. (Aiello et al., 1977; Dul et al., 2011).
Aiello et al. (1977) did research on the effects of workspace crowding over employee's creativity, and they concluded that crowding could have negative effects, regardless of the crowded subjects' interpersonal distance preference, which showed a lower level of creativity than their non-crowded counterparts. Also, Stokols and his team (2002) observed that high levels of environmental distraction, such as noise or prolonged exposure to crowded environments, were associated with less perceived support for creativity at work, and they furthermore suggested that private or non-overcrowded workspaces could have a counter effect, i.e. it could boost employee's creativity.

There has been great progress in terms of gaining a better understanding of the interplay between the work environment and the creativity, or productivity, of office workers. In organizations, employee's creativity can be translated into innovative products, services, processes, systems, work methods, etc. (Dul et al., 2011).

Workplace creativity is usually seen as a result of a creative personality or individual skillsets (Hennessey and Amabile, 2010), dependent on intrinsic motivations, such as personal interest, satisfaction, or the challenge of the work itself. However, there are also studies that suggest that other factors, such as socio-organizational (e.g., job design, team work, rewards, time pressure, and leadership) are also factors in motivating creative work (Campos and Nunes, 2005; Campos et al., 2013; Dul et al., 2011), even in contexts such as kindergarten (Campos and Pessanha, 2011). Many factors contribute to these productivity levels, which are quite subjective. Sometimes it is the room temperature, other times it is the surrounding noise. Even visuals play their role (e.g. if the work desk is messy and cluttered).

### 2.2 Disruptions at Work

In this context, it does not come as a surprise that some researchers have also addressed work interruptions and how technological artifacts could be made in order to reduce those interruptions, e.g. Züger (2017) who propose FlowLight, a device that combines a physical traffic-light-like LED featuring an automatic interruptibility measure based on computer interaction data. As mentioned before, one known socio-organizational factor is crowding and interruptions: crowded environments negatively impact creativity and studies suggest that private workspaces could significantly improve employee creativity. In our approach, it seems interesting to study and gather scientific data about the interplay between interruptions and the overall workspace. In particular, how can we explore the balance between social interruptions and the social benefits of the space. These and other research questions are important for creative industries' workspaces.

## 3 DESIGN APPROACH

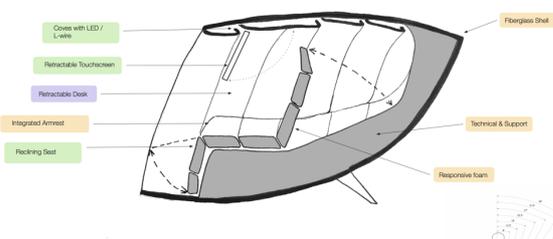

Figure 1: The initial prototypes and sketches for illustrating our multisensorial approach to designing interactive furniture.

Our project takes a multi-disciplinary approach involving architecture, design, and engineering in order to explore how to create a better tool for the workspaces. Our initial prototypes explore the goal of user focusing and the ability for temporary separation through the use of a pod-like chair (See Figure 1). The aim is a design that allows the user to remain in the office, and approachable if necessary while also communicating that they are working on something that requires focus. Sensors and smart technology can detect and respond to the person who is seated, what they are working on (which software they are using, posture, or focus).

Figure 1 also details our design approach in terms of function, technology and form. While the function concerns are relatively straightforward (connected workspace, relaxing, personal), the form concerns we are addressing with the seat prototype include achieving a contemporary, versatile, modular piece of technological furniture that can also bring a timeless, elegant comfort to the workplace. In terms of technology, a reclining seat will provide body positions values that can provide ergonomic feedback to the user. Directional sound provides relaxing soundtracks according to the user's preference, the same applies to the LED interior "mood" lighting.

### 3.1 Pilot Study

Our design approach has the explicit goal of creating interactive furniture to provide a place for relaxation, focus and creativity, something extremely difficult in today's fast-paced offices. To gain an initial understanding of these factors, we conducted a pilot study with participants from a local co-working space, who we invited to take part in an interview, as well as to submit a diary over disruptions in their work over two weeks' time. We distinguish between interruptions and disruptions because of the negative connotation which is typically inherent to interruptions – a disruption can actually be positive for the user, whereas it is never a positive thing to be interrupted. Each of the interviews took 30 minutes and were conducted with two interviewers, one leading the interview and one recording, taking notes, speaking the local language and translating as needed. There were eight participants in the interviews, with four participating in the diary study. The study aimed to explore what they perceived as both positive and negative factors with working in a co-working space. As all participants were currently working in a co-working space, it should be expected that they feel that the positive aspects outweigh the negative. In the interview, they were asked about what they perceived as interrupting their work, but also how they interact with others in the space. In the diary that followed the interview, we can consider comparing how their impressions match up with reality.

The participants focused on two areas, the physical and the social space. The majority of the frustrations expressed was in regards to the physical space, too warm, noisy, uncomfortable, or otherwise not suited for certain tasks, such as calls or meetings, while the spaces dedicated for meetings were not deemed sufficient.

In the discussion of the social space, where the majority of participants expressed that the coworkers were a major positive factor, they also expressed that it was the largest distraction, though not always a negative one.

Joining others for a coffee when struggling with a difficult or tedious task helped maintain focus, although some would have offices that allowed them to close the door when needed. The pilot study suggests that there is a conflict between the public and private in interactions, where the social aspects is at the same time a source of inspiration and motivation, but also interruptions.

### 3.2 Prototyping and Co-creation

In a second phase, we started our prototyping, working together with a team of interior designers, architects, HCI researchers, software and hardware engineers (See Figure 2).

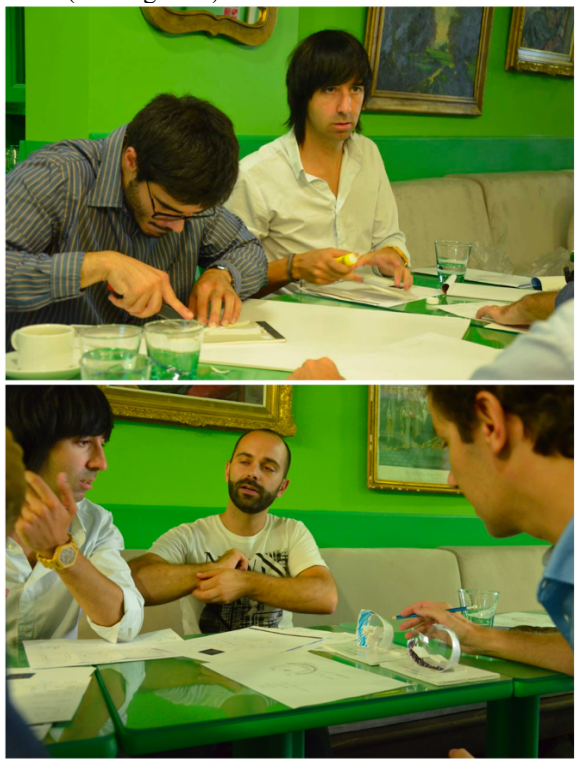

Figure 2: Participatory prototyping session.

We held a 1-day co-creation workshop which employed brainstorming and product ideation techniques with 16 participants, six of them part of the project team, the remaining were creative industries' professionals who had no previous contact or any information about the project.

Figure 2 shows the overall aspect of this co-creation workshop, as well as some of the low-tech prototyping tools and materials which were used. In this workshop, we brought together design professionals, researchers, and developers with the explicit goal of creating concepts together.

We asked them to explore their view on smart furniture through sketching, and low fidelity prototyping. The participants each set out to individually create multiple concepts that were then explored in groups where they combined their ideas, and prototyped them using low-fidelity materials such as paper, foam-core and pencils.

The final concepts which were produced by the different groups of participants involved:

- a Work-life chair, a semi-open pod which can be adjusted for relaxation, lighting, as well as placed into groups for collaboration.
- an all-in-one chair that focuses on the physical com- fort, such as sitting still for too long, or on resting when needed.
- a "sleeper bed" that senses both the environment and the position of the sleeper, but also communicates with its surrounding environment.

The common theme expressed by the participants is relaxation or stress, this suggests that a future design could reflect the need to be able to relax, rather than stressing productivity or focus.

Both the work-life chair and the sleeper bed are intended as parts of a larger environment, rather than as a stand-alone piece.

## 4 TOWARDS MULTI-SENSORIAL APPROACHES

Our position is that this new class of furniture has the overall goal of promoting user focusing and the ability for temporary separation through the use of a pod-like chair (see Figure 3 for 3D renders). The aim is a design that allows the user to remain in the office, and approachable if necessary while also communicating that they are working on something that requires focus. Sensors and smart technology can detect and respond to the person seated, what they are working on (which software they are using, posture, or focus). Although this is ongoing research, it is already possible to draw some conclusions. It becomes noticeable that there are many external factors that can influence one's work productivity. That includes noise and temperature, although the amount of these inputs and its influence can vary from user to user.

Also, human-computer interaction can be used to create better workspaces, either by embedding sensors and actuators in a seat or any other tangible element (e.g. desk).

The human-computer interaction will play its role solely if the technology is embedded in a visual subtle way; that is, in such a way that it contributes positively for the productivity only by giving visual or audio stimulus without distracting the user or adding a lot of complexity to adjust such ambience. It is also interesting to note that visuals play an important role not only in how people work, but also the predisposition that they would have towards using a solution such as the one we have envisioned.

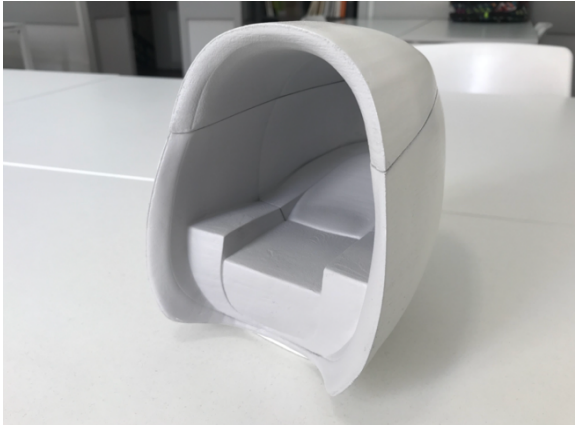

Figure 3: 3D-printed prototype.

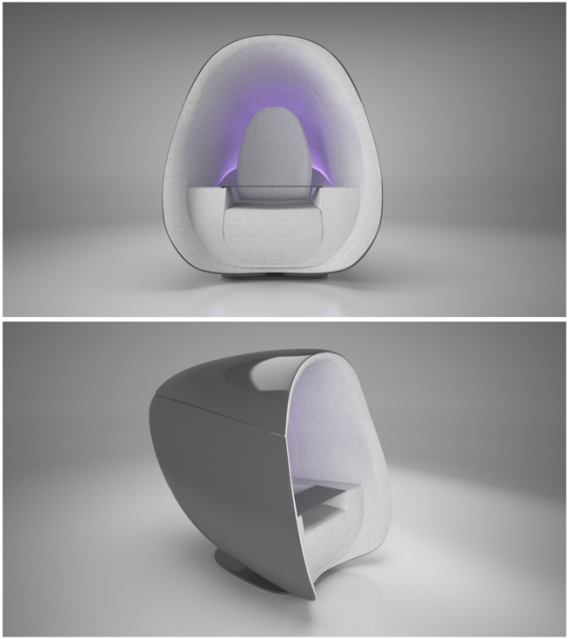

Figure 4: some of the 3D renders of the envisioned interactive seat.

When exploring solutions for the challenges of creative workspaces, we have taken a multi-sensorial approach, in order to explore how different senses, such as hearing or sight can affect productivity. Our design approach is ground in terms of function, technology and form.

While the function concerns are relatively straightforward (connected workspace, relaxing, personal), the form concerns we are addressing include achieving a contemporary, versatile, modular piece of technological furniture that can also bring a timeless, elegant comfort to the workplace. In terms of technology, a reclining seat will provide body positions values that can provide ergonomic feedback to the user. Directional sound provides relaxing soundtracks according to the user's preference, the same applies to the LED interior "mood" lighting.

By sensing the position of the user, we can improve the ergonomics of the chair using adjustable seating, it can also remind the user to take a break through gently adjusting the internal lighting. It helps the user manage interruptions by letting others know externally through outside LED-lighting that a person is working. The aim is to provide simple, effortless interactions.

For the final prototype, we are using pressure-sensitive conductive sheets (Velostat) sewed to woven conductive fabric with a thin conductive thread to build two squares (matrix) for the seat and its back. These allow to track the user's posture and presence.

Afterwards, we added an electret microphone amplifier and an air-quality sensor. Both will be used to check if external factors such as noise and oxygen levels polluted would interfere with the productivity of users.

A temperature and humidity sensor was added so that the system can check the outside air temperature and trigger a ventilation to adjust the inside temperature. In terms of actuators, we added programmable RGB-LED strips to the sides and back of the chair as well as a directional sound beacon just above. Overall, the final prototype will register when a user is sitting down, the ergonomic posture (if posture is correct, the pressure will be distributed along the pressure-sensitive pads), the noise, outside air temperature and air quality (this data would then be cross-checked by the team in terms of minutes against the times when a user would be writing or typing, to see if those had any influence).

Finally, the ambience sound and the LED colouring will be adjusted to test what users would prefer for their two main activities (relax and work). Figure 3 exhibits a 3D print of the prototype, including a cross section that includes a 'cap'. This can be used to allow different configurations of the seat, e.g. relaxing configuration, 'Conference-call' configuration, etc.

This approach brings up several research questions which will be assessed once the prototype becomes a fully usable product: Firstly, it introduces tangible and embedded systems as an approach to resolve the challenge of the physical challenges found at creative workplaces. Secondly, it sheds light on how we can explore the balance between social interruptions and social benefits of the space. Interactive furniture could play a decisive role in the near future (e.g. a smart meeting table could invite other people to join an interesting discussion). Thirdly, it raises some privacy issues: what happens when the office place has a seat like this, where ergonomics data from users is being constantly gathered?

In any workspace, there is a dichotomy between the company and the private. How much time and effort does the employee owe the company, what loyalty? Personal calls during work hours are often frowned upon, but if the call is important the borders soften. If the employer supplies a phone, the employee may receive work email and calls even after hours. While there is no obligation to answer, it is common for employees to respond if they feel it's urgent, or if not, the mere existence causes mental workload.

Therefore, there are many aspects that can be explored; for example, which spaces may be used by anyone? Where is it appropriate to chat and to have coffee? To what extent should one share advice or thoughts with someone from another company?

When creating technology for Smart Workspaces, the intent of various stakeholders shape the needs and requirements for new technology. Small changes in existing designs, sometimes drastically change the nature of the product. Thus, within collaborative workspaces, there's a need to find the right balance between the need for privacy and support for social interaction. As such, collaborative work environments require spaces, furnishings and technologies that facilitate, not only for collaborative work, but also to focus work that fosters solo creativity and productivity.

We are very certain that there are still a lot of issues to explore in this matter, but it seems clear that by combining technology with furniture design and interior design we can actually contribute for improving the worker's productivity and mental well-being.